%% file: ETPathfinder_KIT.tex
\documentclass[a4paper,11pt]{article}
\usepackage{pos}
\usepackage{graphicx}
\usepackage{wrapfig}

\title{The Einstein Telescope Pathfinder and its Vacuum System}

\author[a]{Thomas Höhn}
\author[a]{Adrian Schwenck}
\author[a]{Thomas Thümmler}
\author[b]{Joachim Wolf}
\author[a,b]{Ralph Engel}
\author*[a]{Andreas Haungs}

\author[c]{the Einstein Telescope Pathfinder (ET-PF) collaboration}

\affiliation[a]{Karlsruhe Institute of Technology (KIT), Institute for Astroparticle Physics, D-76021 Karlsruhe, Germany}
\affiliation[b]{Karlsruhe Institute of Technology (KIT), Institute for Elementary Particle Physics, D-76021 Karlsruhe, Germany}
\affiliation[c]{Partners of ET-PF see \url{https://www.etpathfinder.eu/partners/}; a complete list of authors can be found at the end of the proceedings}

\emailAdd{andreas.haungs@kit.edu}

\abstract{
The Einstein Telescope (ET) will be the next generation gravitational wave observatory in Europe with a sensitivity
reaching beyond the CMB into the dark era of the Universe.
Each corner of the triangular baseline
design is the center of two interferometers with 10\,km long arms, one operated at room temperature, the other
one with mirrors at cryogenic temperatures of 10\,–\,15 K that reduce the noise contribution at frequencies as
low as 3\,Hz.
The ETpathfinder (ET-PF) project at Maastricht University is a R\&D facility for the challenging cryogenic interferometer
technology of ET. It is a 20\,m~x~20\,m interferometer with six towers that will house the seismically
decoupled cryogenic Si-mirrors, laser systems, and detectors. The KIT group developed the control system of
the ultra-high vacuum system for ET-PF, based on the expertise from the KATRIN neutrino mass experiment. In
addition, a test facility is currently being set up at KIT to investigate adsorption and desorption processes of
residual gas on the cryogenic mirror surfaces, as well as monitoring techniques and in-situ cleaning procedures.
This paper presents the objectives and status of these activities and their contribution towards the next
generation gravitational wave observatory.
}

\ConferenceLogo{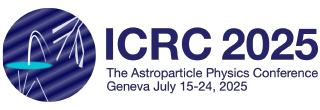}

\FullConference{39th International Cosmic Ray Conference (ICRC2025)\\
 15–24 July 2025\\
Geneva, Switzerland\\}

\begin{document}
\maketitle


\section{The Einstein Telescope – A Next-Generation Gravitational Wave Observatory}

The Einstein Telescope (ET) is the proposed next-generation underground observatory for gravitational waves~\cite{lit:ET,lit:ET1,lit:ET2} in Europe, designed to significantly surpass the sensitivity of current detectors, such as LIGO~\cite{lit:LIGO}, Virgo~\cite{lit:VIRGO} or KAGRA~\cite{lit:KAGRA}. It will enable scientists to explore the universe through gravitational waves with unprecedented precision and reach, opening new windows into the early universe, black hole mergers, neutron star collisions, and other fundamental phenomena~\cite{lit:ET2}.

\begin{wrapfigure}[]{r}[0.1cm]{0.5\textwidth}
\includegraphics[width=0.48\textwidth]{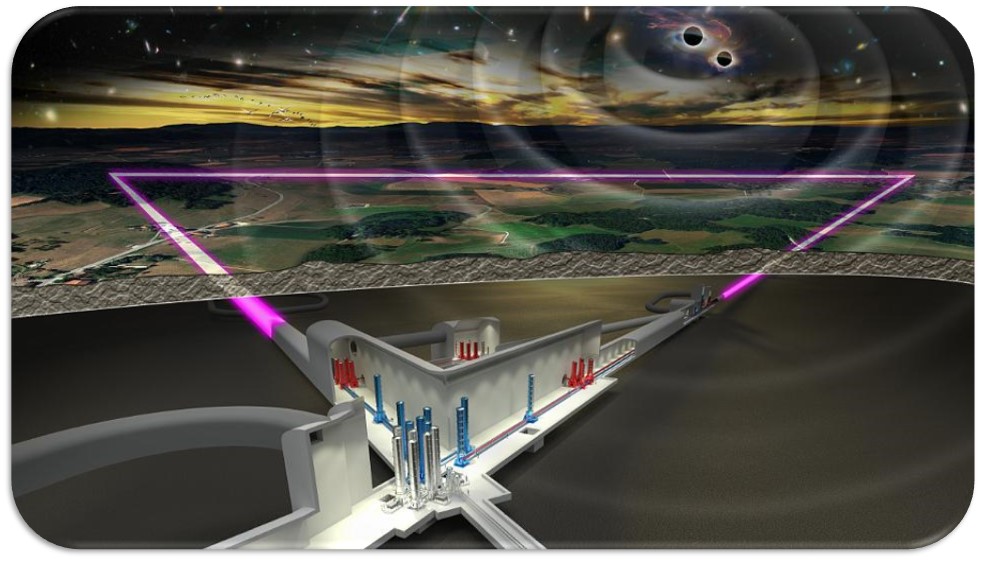} \\
\boxed{\includegraphics[width=0.46\textwidth]{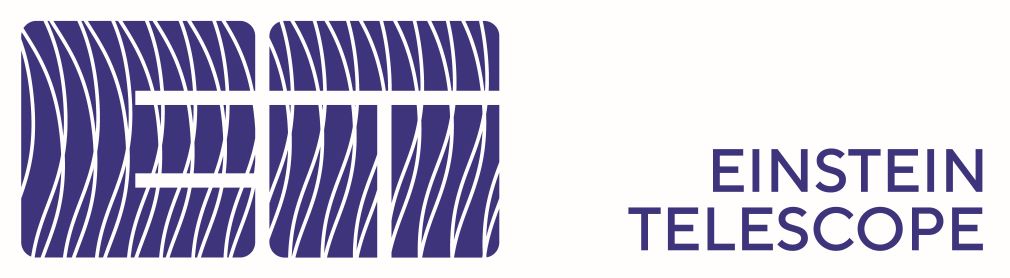} }
	\protect \caption{Top: Artistic sketch of the Einstein Telescope in the triangle configuration to be installed underground (credit Nikhef). Bottom: Logo of the Einstein Telescope project.} 
    \label{fig:ET}
\end{wrapfigure}
One of the proposed sites for the Einstein Telescope is located in the EMR (Euregio Meuse-Rhine) region, at the intersection of Belgium, Germany, and the Netherlands, and another one in Sardinia, Italy. 
Recently, a third candidate for the site from the Lusetia region in Saxony, Germany was accepted. 
The baseline design consists of a triangular layout with 10\,km long arms, forming a 30\,km underground network. Each corner is the center of two interferometers, one operating with mirrors at room temperature (RT) and the other at cryogenic temperatures (CT) to maximize sensitivity across a wide frequency range. This design accommodates four ultra-high vacuum interferometer tubes with a diameter of 1\,m in each arm of the triangle. 
An alternative scenario is discussed in which an L-shaped tunnel with 15\,km long arms is installed at two different locations, each with interferometers in both temperature regimes.

In order to achieve the desired sensitivity, ultra-high vacuum (UHV) conditions at a pressure as low as $10^{-10}\,$mbar have to be maintained in the complex system of more than 120\,km of vacuum tubes and approximately 180 towers for the optical equipment with a combined volume of about 100\,000\,m$^3$ (triangle configuration).
These extreme vacuum conditions are necessary to isolate the interferometers from environmental noise and prevent fluctuations in the refractive index in air, to allow the detection of the faintest space-time distortions. Good vacuum conditions are also required as thermal insulation of the cryogenic mirrors and to mitigate adsorption beyond a monolayer (ML) of residual gas on the cold mirror surfaces.

As a European flagship project in fundamental physics, the Einstein Telescope represents a major step toward understanding the gravitational universe, advancing technology, and fostering international scientific collaboration.

\section{ETpathfinder: Preparing the Ground for the Einstein Telescope}

The Einstein Telescope Pathfinder (ET-PF) is a crucial research and development facility~\cite{Utina:2022qqb,lit:ETPF} dedicated to advancing the technologies required for the Einstein Telescope. Located in Maastricht, ET-PF provides a state-of-the-art, low-noise environment for prototyping and testing key components of future gravitational wave detectors.

\begin{figure}[!t]
    \centering
	\includegraphics[width=1\textwidth]{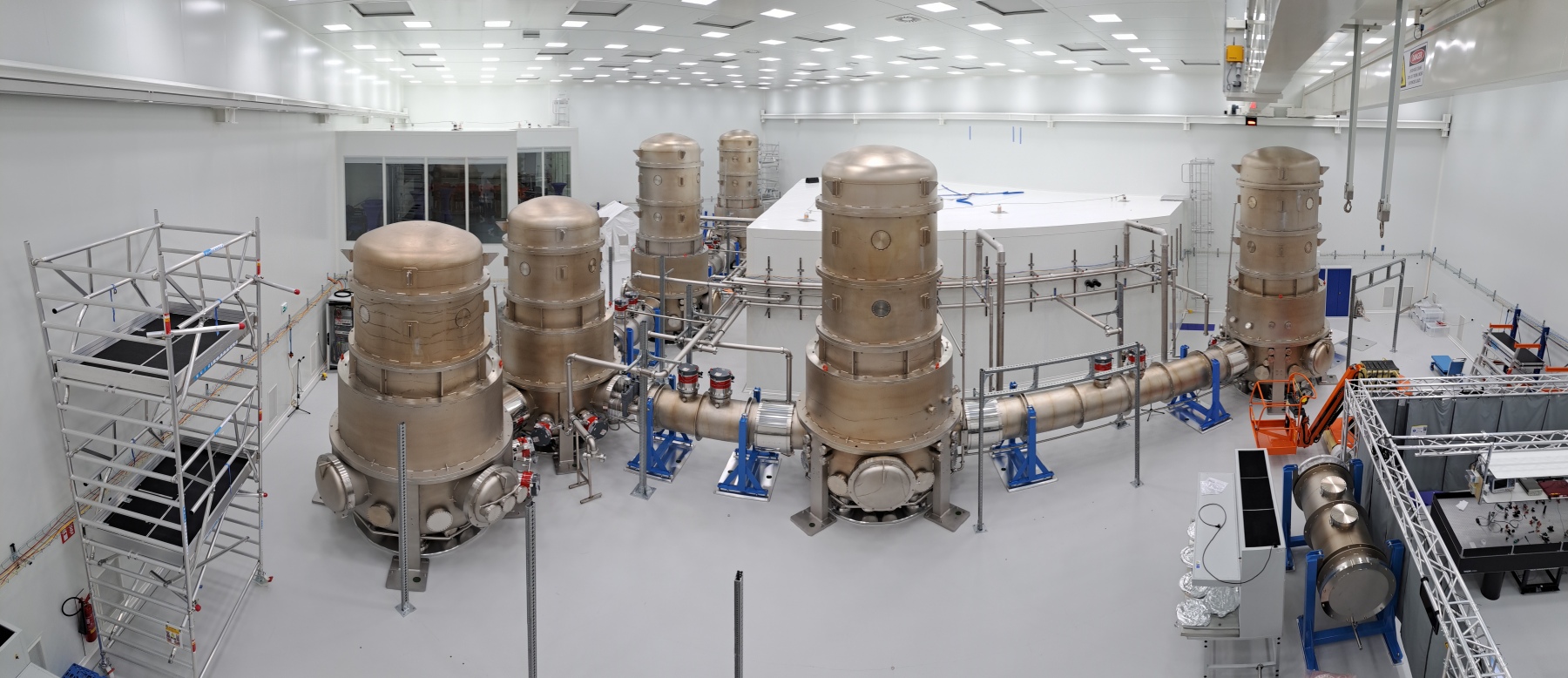}
	\protect\caption{Photo of the ET-PF installation in Maastricht, NL.}
    \label{fig:ET-PF}
\end{figure}
The Einstein Telescope is designed to deliver a ten-fold sensitivity improvement in the mid and high-frequency range of gravitational waves. However, to access the low-frequency band below $10\,$Hz, an even greater leap is required: improvements by factors of 100 to 1000 over current capabilities. Achieving this dramatic enhancement poses significant scientific and technological challenges, especially in terms of vibration isolation, thermal noise control, and quantum measurement techniques.

ET-PF addresses these challenges by hosting a full interferometer setup under controlled conditions, enabling researchers to develop, integrate, and refine the novel technologies needed for low-frequency detection. This includes cryogenic systems, ultra-stable lasers, seismic isolation platforms, and quantum-enhanced readout schemes.

More than 20 institutions and research groups from across Europe - including the Netherlands, Belgium, Germany, France, Poland, Spain, and the UK - are collaborating on ETpathfinder. This strong international partnership not only advances fundamental physics, but also strengthens Europe’s role in pioneering gravitational wave science and precision engineering.

As a stepping stone to the Einstein Telescope, ETpathfinder plays a vital role in securing the success of the future observatory. It provides a unique environment where ideas can be turned into working technologies, bridging the gap between theoretical designs and real-world implementation.

\subsection{KIT Contribution to ETpathfinder: Technology Transfer from KATRIN}

The Institute for Astroparticle Physics at KIT contributes to the Vacuum \& Cryogenics Task of the ET-PF with expertise gained from the KATRIN neutrino experiment~\cite{lit:KATRIN_web}. At KATRIN, KIT designed, built, and successfully operates one of the world’s most advanced ultra-high vacuum (UHV) systems, maintained reliably over more than a decade.

This unique experience is now being transferred to ET-PF. KIT leads the design of the ET-PF vacuum control system, based on Siemens PCS7~\cite{lit:pcs7}, including the specification of circuit diagrams, hardware configurations, and interlock systems for safe and robust long-term operation.

Special attention is given to the integration of the vacuum and cryogenic systems and their seamless interaction with other ET-PF subsystems. This contribution ensures that the demanding environmental conditions for gravitational wave detection — particularly in the low-frequency range — can be met with proven technological solutions adapted from KATRIN~\cite{lit:KATRIN}.

In addition, available vacuum setups from KATRIN at KIT have been modified to be used for the investigation of gas adsorption and desorption on cold surfaces, such as silicon, which is the material of choice for the RT and CT mirrors. Materials can be tested in a dedicated outgassing system before being installed in the UHV-chamber for further measurements. 

\section{Vacuum and Cryogenic System}

The ultra-high vacuum (UHV) and cryogenic systems of ETpathfinder~\cite{lit:ETPF-design}
are designed to facilitate the development and demonstration of techniques that can be used in future third-generation cryogenic gravitational wave interferometers. The Michelson interferometer includes the input and end mirrors of the Fabry-Perot cavities in the two arms, which are cooled down to cryogenic temperatures to reduce fundamental thermal noise. 
The design aims to reduce the level of vibrational noise on the mirror surfaces below $10^{-18}\,{\rm m/\sqrt{Hz}}$ at 10\,Hz. The mirrors are planned to be operated at RT, 123\,K and 10\,K. Both the cryogenic system and the vacuum system should not limit the noise budget in any way. 

The KIT group developed the vacuum control system based on the experience gathered by the long-term operation of the KATRIN experiment \cite{lit:KATRIN} using the Siemens PCS7 process control system.
The cryogenic system has been designed and installed by industrial partners and its control system is provided by company Demcon Kryoz.

\begin{figure}[ht]
    \centering
	\includegraphics[width=1\textwidth]{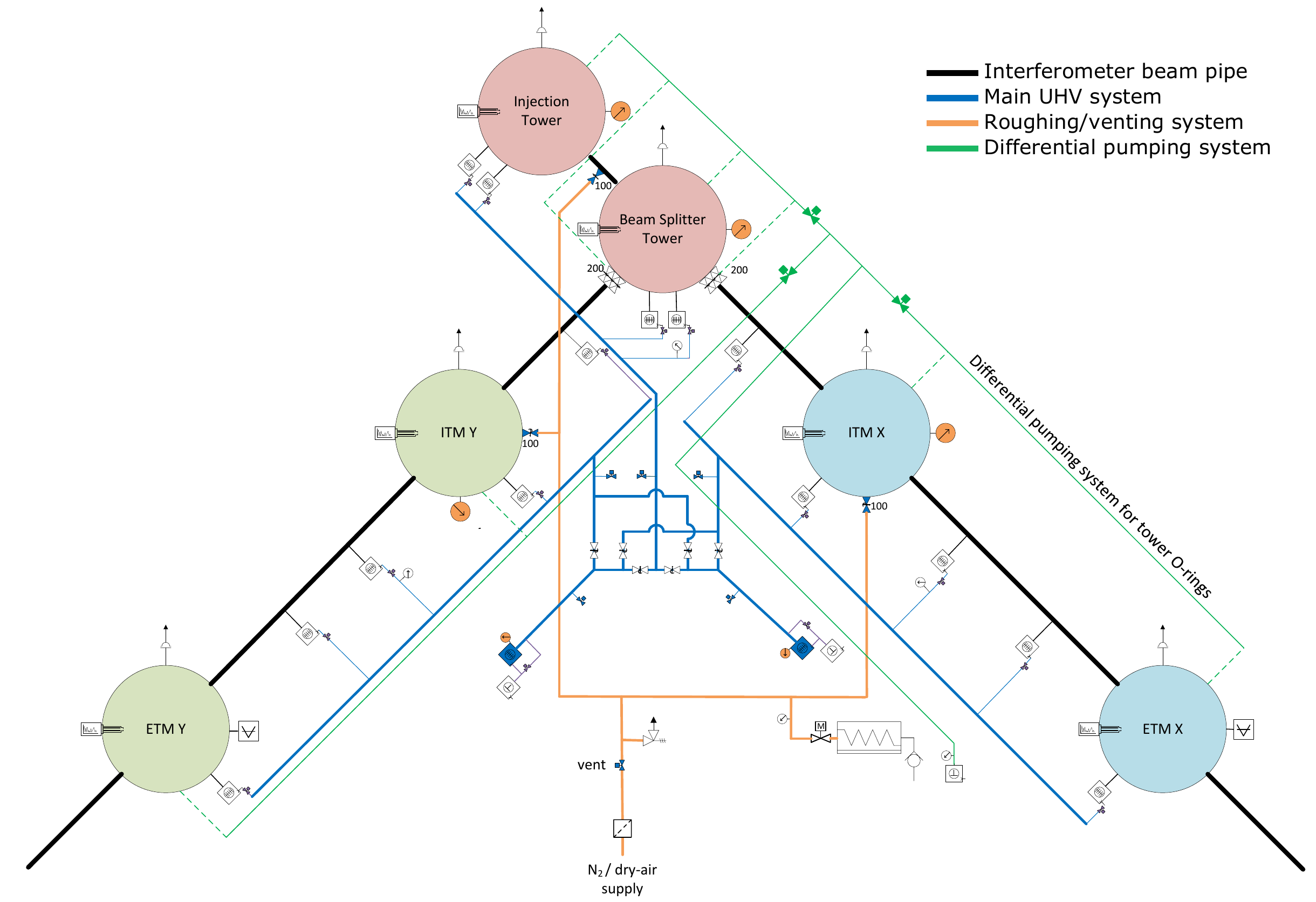}
	\protect\caption{P\&I diagram of the ETpathfinder vacuum system. The six towers are connected by pipes (black) housing the actual interferometer. A central part with injection and beam splitter tower. North (Y) and south (X) arms with two mirror towers each. Main ultra-high vacuum system (blue), roughing and venting system (orange), and differential pumping for the large flanges of the towers (green).}
    \label{fig:ET-PF-PID}
\end{figure}

The baseline design of the ETpathfinder setup is L-shaped. In the central part, two vacuum towers are installed for injection and beam splitting. From the beam splitter tower two arms, hosting the Fabry-Perot cavities, point into north west (Y arm) and south west (X arm) directions,  each equipped with two vacuum towers.
The towers are 6.1\,m high with a diameter of 2.8\,m, and connected by beam pipes of 800\,mm diameter. 
The Fabry-Perot cavity has a length of 9.3\,m, however the complete length from injection along the X arm adds up to 25\,m.
In total the vacuum system has to provide UHV conditions in a volume of about 170\,m$^3$.
The P\&ID of the setup is shown in Fig.~\ref{fig:ET-PF-PID}. More details can be found in the design report~\cite{lit:ETPF-design}.

In order to provide the required pumping speed, a cascaded pumping system has been set up. In total 14 large DN 320 turbo-molecular pumps with up to 3050\,l/s pumping speed (Pfeiffer ATH 3204) are installed. Thereof two pumps at each tower in the central section and one pump at each other tower. The remaining pumps are installed directly at the beam tube.
The three separate sections are connected by DN 100 pipes to a 2-fold fore-vacuum system consisting of a DN 100 turbo-molecular pump (Pfeffer HiPace 350) backed by a multi roots pump.
This system provides flexibility and redundancy during operation and testing.

A separate roughing vacuum system is installed to quickly pump down the towers and beam tubes before the UHV system can take over. 
Using separate lines directly connected to the main volumes of each section, a dry screw pump (Pfeiffer Hepta 630P) has sufficient capacity to pump down from atmospheric pressure into a regime where the UHV system can be activated.

Finally, a third independent pumping system with just one multi roots pump provides the differential vacuum for the space between the large double O-rings of the towers.
This is necessary to limit the permeation at the large tower flanges.

In order to provide safe and reliable operation, all parts of the vacuum system have been collected into one comprehensive circuit diagram, covering all power and data signal needs of the individual components.
One central cabinet, housing the CPU of the Siemens PCS7 system, is connected to three more cabinets next to each section of the vacuum setup. 
The cabinets and circuit layouts have been designed and built according to industry standard.

Using separate data buses for hardware operation, data exchange, and operator intervention, the system provides a high level of reliability and intrinsic safety.
The PCS7 CPU talks directly to a pair of redundant process servers, but is self-sustaining by its own. Even when loosing all communication, the vacuum system will continue to run and all safety measures and interlocks are effective.

Those process servers communicate to a network of client stations in the control room and next to the hardware inside the clean room.
Furthermore, process data is exchanged with the central database of the experiment as well as with the cryogenic system for safety reasons.

The PCS7 control software is based on an in-house developed library of function modules originating from the KATRIN experiment, but modified and extended for ET-PF.
Boundary conditions and interlocks are implemented depending on the actual operation mode like pumping down, venting, or standard operation. 
For each of those modes the safety levels may differ, parts need to be protected or just restricted.

Based on the long-term reliable operation of the KATRIN experiment, we are confident that this system will provide safe and steady operation of the Einstein Telescope Pathfinder in order to facilitate uninterrupted R\&D for the next generation of gravitational wave observatories. 

\section{Adsorption and Desorption at Cryogenic Mirrors}

The requirements for the vacuum condition for the cryogenic mirrors (10\,K) are even stronger than those for RT mirrors \cite{lit:cryo-design}. Residual gas molecules can cryosorb on the cold surfaces, accumulating monolayer by monolayer on the surface. Simulations indicate a pressure below $10^{-12}\,$mbar with a partial pressure of $10^{-14}\,$mbar for water in ET. Under these optimum conditions, water ice would accumulate at a rate of one monolayer (ML) per year. Each additional ML adds to the heat load of the cooling system, requiring a strong limit on the maximum layer thickness of only a few ML. 

\begin{figure}[ht]
    \centering
	\includegraphics[width=0.85\textwidth]{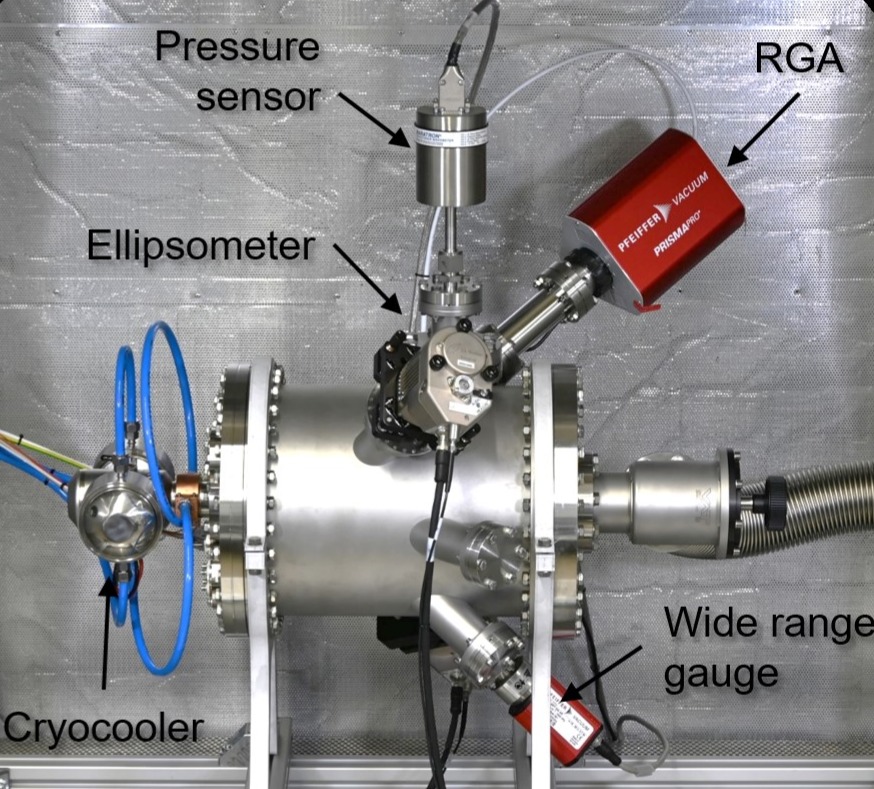}
	\protect\caption{'Vacuum, surface monItoring aNd Cleaning experiment for EinsteiN Telescope' (VINCENT): UHV vacuum chamber for adsorption tests with RGA, wide-rage BA vacuum gauge, Baratron capacitance gauge, ellipsometer, cryo cooler and pumping system.}
    \label{fig:Ellipsometer}
\end{figure}
At KIT the 'Vacuum, surface monItoring aNd Cleaning experiment for EinsteiN Telescope' (VINCENT) has been set up, mainly consisting of a versatile UHV test chamber that will allow detailed investigations of the adsorption of gas at different surface temperatures from room temperature to 40\,K. 
The test stainless-steel chamber, shown in Fig.~\ref{fig:Ellipsometer}, has a diameter of 250\,mm and a similar length. It is equipped with a residual gas analyser (RGA) for partial pressure measurement, to see what type of gases will be adsorbed and desorbed on a cold Si-surface that represents the mirrors. The absolute pressure can be measured with a wide-range hot cathode vacuum gauge and a gas-type independent capacitance gauge (Baratron). The vacuum is produced by a turbo-molecular pump backed by a scroll pump. During initial tests without backing, the system reached $10^{-8}$\,mbar or lower. This will be further optimized by  bakeout at moderate temperatures. In order to cool the mirror, a sterling-type Lihan TC4188 cryocooler with a cooling power of 15\,W at 77\,K or 2\,W at 40\,K is used in the first step. In addition, a thermal radiation shield is installed around the mirror and the mirror mount to reduce heat transfer via thermal radiation.

The thickness of the accumulating gas layer on the cold surfaces will initially be monitored by ellipsometry for different gas species. A second method to monitor the growing thickness of the layer will use a quartz crystal microbalance (QCM). It will be cooled down to the same temperature as the Si sample without touching it. This would have the advantage of monitoring the accumulating gases without disturbing the delicate stability of the mirror suspension of ET. However, the frequency change of the QCM with growing layer mass depends strongly on the temperature. Therefore, it is planned to cross-calibrate the QCM with the ellipsometry system and determine the temperature relation. 

Additional measurements planned with the system include the test of in-situ cleaning methods of the cold surface, for instance with low energy electrons, UV light, or Ar plasma.   

\section{Conclusion and Outlook}

The third generation European gravitational wave observatory, the Einstein Telescope, requires a strong R\&D effort to reach the ambitious goal to surpass the sensitivity of existing observatories by another order of magnitude. 

The Einstein Telescope Pathfinder was set up to provide a realistic environment for R\&D, for testing under real conditions and for learning and training scenarios for the integration of specific parts for the Einstein Telescope under quasi-real conditions.

KIT started to work on various aspects of the vacuum and cryogenic requirements. Experience with huge UHV systems, such as the KATRIN experiment, will aid to reach this goal. Contributions to the success of the ETpathfinder R\&D facility in Maastricht, as well as dedicated test setups at KIT to understand, monitor, and control gas adsorption on cold mirror surfaces will add to achieving the final goal of a unique European contribution to "listen" to and understand the early universe.   

\newpage

\paragraph{Acknowledgements:}
The authors would like to thank the entire ET-PF team as well as the members of the KATRIN Collaboration for the constructive cooperation and the possibility of technology transfer in order to advance and realise the major goal of a future Einstein Telescope. 


\clearpage


\input{authorlist_ETPF.tex}

\end{document}

%% file: authorlist_ETPF.tex
\section*{\textcolor{darkgray}{Full Author List (July, 2025):} \\ 
Einstein Telescope Pathfinder Collaboration (ET-PF)\\
}

\footnotesize
\noindent
M.~Adams$^2$, 
K.~Akhil$^7$,  
A.~Amato$^{1,2}$, 
M.~Baars$^2$, 
P.~Baer$^6$, 
W.~Beaumont$^7$, 
C.~Bellani$^8$, 
A.~Bertolini$^2$, 
S.~Biersteker$^3$, 
A.~Binetti$^8$
% ...
\\
\\
\noindent
 $^1$ Maastricht University, Department of Gravitational Waves and Fundamental Physics, 6200 MD Maastricht, Netherlands  \\
 $^2$ Nikhef, Science Park 105, 1098 XG Amsterdam, Netherlands  \\
 $^3$ Vrije Universiteit Amsterdam, Department of Physics and Astronomy, De Boelelaan 1085, NL-1081 HV Amsterdam, Netherlands  \\
% ...
$^6$ Fraunhofer ILT - Institute for Laser Technology, Steinbachstr. 15, 52074 Aachen, Germany  \\
 $^7$ Universiteit Antwerpen, Prinsstraat 13, 2000 Antwerpen, Belgium  \\
 $^8$ KU Leuven (Department of Physics and Astronomy, Celestijnenlaan 200D, 3001 Leuven, Belgium  \\
% ...

\subsection*{Acknowledgments}

\noindent
The ETpathfinder project in Maastricht is funded by Interreg Vlaanderen-Nederland, the province of Dutch Limburg, the province of Antwerp, the Flemish Government, the province of North Brabant, the Smart Hub Flemish Brabant, the Dutch Ministry of Economic Affairs, the Dutch Ministry of Education, Culture and Science, and by own funding of the involved partners. 
In addition the ETpathfinder team acknowledges support from the European Research Council (ERC), the Dutch Research Council (NWO), the Research Foundation Flanders (FWO), the German Research Foundation (DFG), 
Spanish MICINN, the CERCA program of the Generalitat de Catalunya and the Dutch National Growth Fund (NGF).